\documentclass[twocolumn]{aastex62}
\usepackage{placeins}
\begin{document}
\title{Chlorine Isotope Ratios in M Giants}

\author{Z. G. Maas}
\author{C. A. Pilachowski}
\affil{Indiana University Bloomington, Astronomy Department, 727 East Third Street, Bloomington, IN 47405, 
USA}
\email{zmaas@indiana.edu}

\begin{abstract}

We have measured the chlorine isotope ratio in six M giant stars using HCl 1-0 P8 features at 3.7 microns with R $\sim$ 50,000 spectra from Phoenix on Gemini South. The average Cl isotope ratio for our sample of stars is 2.66 $\pm$ 0.58 and the range of measured Cl isotope ratios is 1.76 $<$ $^{35}$Cl/$^{37}$Cl $<$ 3.42. The solar system meteoric Cl isotope ratio of 3.13 is consistent with the range seen in the six stars. We suspect the large variations in Cl isotope ratio are intrinsic to the stars in our sample given the uncertainties. Our average isotopic ratio is higher than the value of 1.80 for the solar neighborhood at solar metallicity predicted by galactic chemical evolution models. Finally the stellar isotope ratios in our sample are similar to those measured in the interstellar medium. 

\end{abstract}


\section{Introduction}
\label{sec::intro}

The odd, light elements are useful for understanding the production sites of secondary nucleosynthesis processes. However, some of the odd light elements, such as P, Cl, and K have few measured stellar abundances and/or do not match predicted chemical evolution models (see \citealt{nomoto13} for a review). For example, chlorine may be made through multiple nucleosynthesis processes but few studies of Cl in the Galaxy exist.

Cl has two stable isotopes. $^{35}$Cl is thought to be produced primarily by proton capture on $^{34}$S during explosive oxygen burning ($^{34}$S(p,$\gamma$)), where free protons are created from the $^{16}$O + $^{16}$O reaction or from photodisintigration \citep{woosley73}. $^{37}$Cl is thought to be produced primarily by the decay of $^{37}$Ar (produced via neutron capture on $^{36}$Ar) during oxygen burning in core collapse supernova \citep{woosley73,thielemann85,woosley95}. 

In core collapse supernova (CCSNe), the mass and metallicity of the supernova can impact the isotopic ratio of Cl. Examples of yields from different core CCSNe models are listed in Table \ref{table::yields}. The weak s-process in massive stars may also be a significant source of $^{37}$Cl. Models predict that $^{37}$Cl production increases with He-core mass and neutron excess \citep{prantzos90}. For example, in a 25 M$_{\odot}$ solar mass star, $^{37}$Cl can be over-abundant by a factor of nearly 50 compared to the solar system $^{37}$Cl abundance \citep{pignatari10}. $^{37}$Cl production via the s-process in AGB stars is not thought to be as significant as from the weak s-process \citep{cristallo15,karakas16}. For example, FRUITY models predict only a $\sim$ 3$\%$ increase in $^{37}$Cl for a 4 M$_{\odot}$ solar metallicity AGB star and a 14$\%$ increase from the initial surface abundance to that after the final dredge up for a 2 M$_{\odot}$, solar metallicity AGB star \citep{cristallo15}. Also, \citet{karakas16} predict a 3  M$_{\odot}$ stars with Z = 0.014 and an initial Cl isotope ratio of 3.13, will end with an isotope ratio of $\sim$ 2.6 at the tip of the asymptotic giant branch.

A small amount of chlorine is also predicted to be created during Type Ia supernovae with an isotope ratio between 3 - 5 depending on the model parameters \citep{travaglio04,leung17}. However, Type Ia supernovae yields are not as significant as CCSNe, since the explosive material has little hydrogen available for proton capture \citep{leung17}. For example, the benchmark models of \citet{travaglio04} and \citet{kobayashi11} demonstrate yields from Type Ia supernovae are an order of magnitude smaller than CCSNe yields, as shown in Table \ref{table::yields}. Finally, $^{35}$Cl may also be produced from neutrino spallation during CCSNe \citep{pignatari16}. A summary of the different yields and production factors from these sources are listed in Table \ref{table::yields}.

The chemical evolution model from \citet{kobayashi11} predicts Cl isotope ratios in the solar neighborhood of $^{35}$Cl/$^{37}$Cl = 1.94 at [Fe/H] = --0.5 and a Cl isotope ratio of $^{35}$Cl/$^{37}$Cl = 1.80 at solar metallicity. This value is lower than the solar system meteoric Cl abundance of 3.13 \citep{lodders09}. 

Cl abundance measurements are difficult in stellar spectra due to a low abundance and no strong optical absorption features. $^{35}$Cl abundances in stars have been measured using the H$^{35}$Cl feature located at 3.7 $\mu$m in stars with T $<$ 4000 K \citep{maas16}. Stars with temperatures above $\sim$ 4000 K do not have HCl features in their spectra due to the molecule's low dissociation energy. In the star RZ Ari, both the H$^{35}$Cl and H$^{37}$Cl were detected and a Cl isotope ratio of 2.2 $\pm$ 0.4 was derived \citep{maas16}. This star is the coolest of their sample with an effective temperature of 3340 K \citep{mcdonald12}. 

\begin{deluxetable*}{ llcccccc}
%\tabletypesize{10pt} 
%\rotate
\tablewidth{0pt} 
\tabletypesize{\scriptsize}
\tablecaption{Nucleosynthesis Predictions \label{table::yields}} 
 \tablehead{\colhead{Nucleosynthesis Site} & \colhead{Model Parameters} & \colhead{source} & \colhead{Yields $^{35}$Cl} & \colhead{Yields $^{37}$Cl} & \colhead{Production} & \colhead{Production} & \colhead{$^{35}$Cl/$^{37}$Cl\tablenotemark{a}}\\
 \colhead{} & \colhead{} & \colhead{} & \colhead{(M$_{\odot}$)} & \colhead{(M$_{\odot}$)} & \colhead{Factor $^{35}$Cl} & \colhead{Factor $^{37}$Cl} & \colhead{}}
\startdata
CCSNe &	13 M$_{\odot}$, Z = 0.02, E = 10$^{51}$ ergs & 1 & 1.15 x 10$^{-4}$ & 3.03 x 10$^{-5}$ & \nodata & \nodata & 3.80 \\
CCSNe & 13 M$_{\odot}$, Z = 0.014 & 2 & 1.18 x 10$^{-4}$ & 2.79 x 10$^{-5}$ & \nodata & \nodata & 4.23 \\
CCSNe & 13 M$_{\odot}$, Z = 0.014, rotation included& 2 & 1.64 x 10$^{-4}$ & 4.82 x 10$^{-5}$ & \nodata & \nodata & 3.40 \\
CCSNe &	25 M$_{\odot}$, Z = 0.02, E = 10$^{51}$ ergs & 3 & 3.63 x 10$^{-4}$ & 3.03 x 10$^{-4}$ & \nodata & \nodata & 1.20 \\
CCSNe & 25 M$_{\odot}$, Z = 0.014 & 2 & 2.92 x 10$^{-4}$ & 1.25 x 10$^{-4}$ & \nodata & \nodata & 2.34 \\
CCSNe & 25 M$_{\odot}$, Z = 0.014, rotation included & 2 & 5.45 x 10$^{-4}$ & 1.85 x 10$^{-4}$ & \nodata & \nodata & 2.95 \\
\hline
Type Ia SNe & Model b30 3d 768\tablenotemark{b} & 4 & 4.58 x 10$^{-5}$ & 1.21 x 10$^{-5}$ & \nodata & \nodata &  3.79 \\
\hline
AGB Star\tablenotemark{c} & 3 M$_{\odot}$, Z = 0.014 & 5 & \nodata  & \nodata & 0.90 & 1.42 & 2.34 \\
AGB Star\tablenotemark{d} &  3 M$_{\odot}$, Z = 0.014 & 6 & \nodata & \nodata & 0.997 & 1.13 & 2.60 \\
\hline
Weak s-process in He and C Shells& 25 M$_{\odot}$, Z = 0.014 & 7 & \nodata & \nodata & 0.2 - 0.3 & 46 - 47 & \nodata \\
\enddata
\tablenotetext{a}{Solar System Meteoric $^{35}$Cl/$^{37}$Cl = 3.13 \citep{lodders09}}
\tablenotetext{b}{Benchmark model}
\tablenotetext{c}{Initial, pre-AGB evolution $^{35}$Cl/$^{37}$Cl = 3.13. Net yield for $^{37}$Cl = 5.89 x 10$^{-7}$ $M_{\odot}$}
\tablenotetext{d}{Initial, pre-AGB evolution $^{35}$Cl/$^{37}$Cl = 2.94. Net yield for $^{37}$Cl = 1.60 x 10$^{-7}$ $M_{\odot}$}
\tablecomments{Sources: (1) \citealt{kobayashi06}; (2) \citealt{chieffi13}; (3) \citealt{kobayashi11}; (4) \citealt{travaglio04}; (5) \citealt{karakas16}; (6) \citealt{cristallo15}; (7) \citealt{pignatari10}}
\end{deluxetable*}

The Cl isotopic ratio has been explored in the interstellar medium (ISM) using chlorine bearing molecules in the millimeter/radio regime. Surveys of the Cl isotope ratio have used HCl features and have found a range of isotope ratios in the ISM, from 1 $\lesssim$ $^{35}$Cl/$^{37}$Cl $\lesssim$ 5. HCl in the interstellar medium has been examined using both space based and ground based observatories. Chloronium (H$_{2}$Cl$^{+}$) has also been used to probe Cl isotope ratios after the molecule was discovered in the interstellar medium using the Herschel Space Observatory \citep{lis09}. Finally, the Cl isotope ratio has been derived in the circumstellar envelopes of evolved stars using NaCl, AlCl, KCl, and HCl. A discussion of ISM  $^{35}$Cl/$^{37}$Cl measurements can be found in section \ref{subsec::ism} (see also Table \ref{table::ism_ratio}).

The range of measured isotope ratios may reflect different nucleosynthesis histories of the material but the systematic errors that may have been introduced in different studies make comparisons difficult. Also, chemical fractionation is not expected to be significant for Cl due to the similar masses of each isotope \citep{kama15}. To test predictions of Cl nucleosynthesis, we have measured the Cl isotope ratio in M giants. Observations and data reductions are discussed in section \ref{sec::obs}. The methodology used to derive Cl isotope ratios is discussed in section \ref{sec::methodology}. A discussion of the results can be found in section \ref{sec::discussion}. We summarize our conclusions in section \ref{sec::conclusion}

\section{Observations and Data Reduction}
\label{sec::obs}

\begin{figure*}[t!]
	\centering 
 	\includegraphics[trim=0cm 0cm 0cm 0cm, scale=.22]{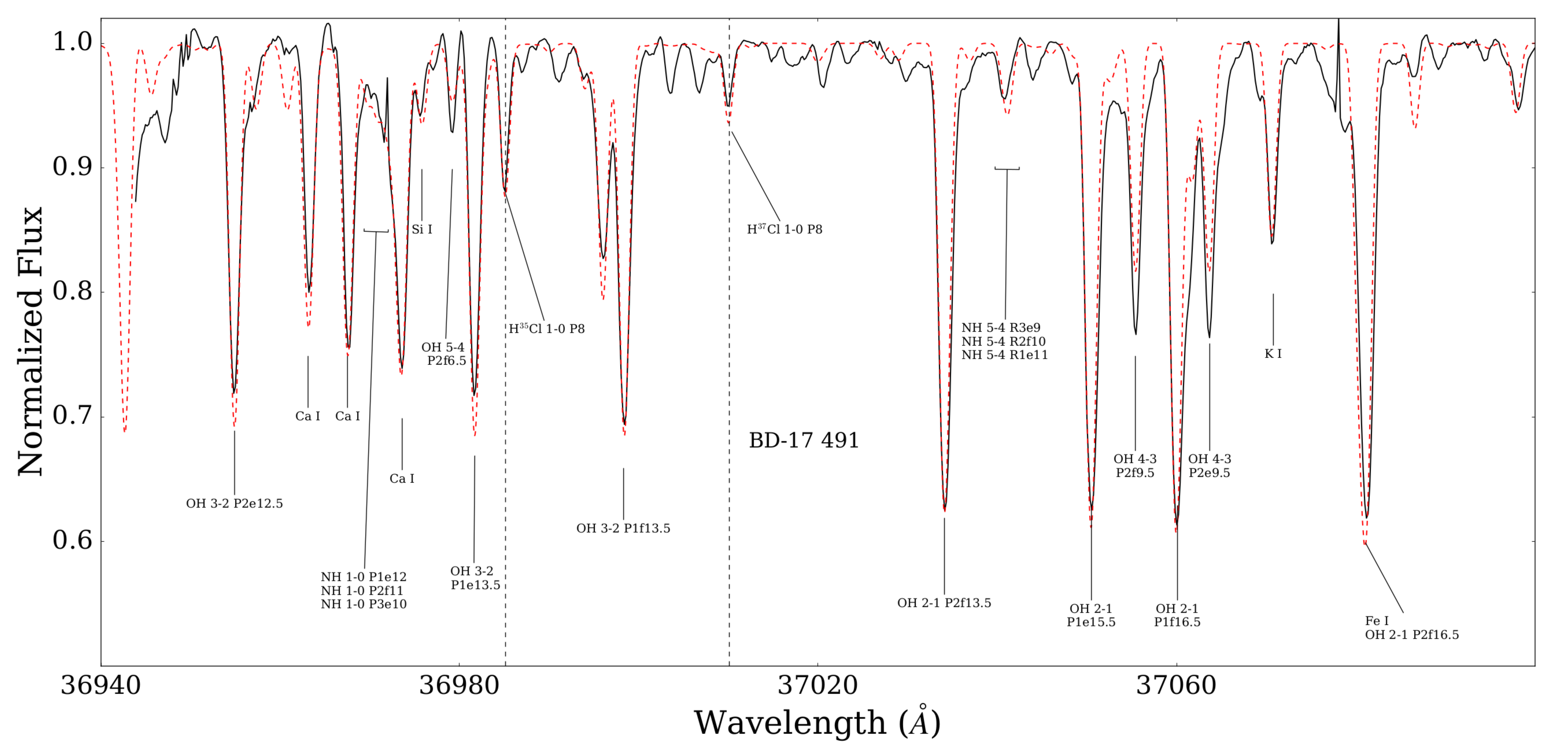}
	\caption{Spectrum of the M-star BD -17 491. The red dashed line represents a model spectrum created to fit the absorption lines for this star. Line identifications are added with the sources for each transition found in section \ref{subsec::synthesis}. \label{fig:spectra} }
	\end{figure*}

We chose stars from the 2MASS and WISE catalogs \citep{skrutskie06,wright10}. Only cool stars are expected to have both the H$^{35}$Cl and H$^{37}$Cl features due to the low dissociation energy of the HCl molecule. To ensure our sample of stars contained both HCl features, we calculated the effective temperatures of our stars (as described in section \ref{subsec:params}), and chose stars with photometric temperatures similar to RZ Ari at $\sim$ 3340 K. RZ Ari was the  coolest star in the sample of \citet{maas16} and the only star in which H$^{37}$Cl was detected. Stars redder than RZ Ari, with J-K$_{s} \gtrsim $ 1.26, were initially selected as target stars. We selected stars between 2.0 mags $<$ K$_{s}<$ 3.0 mags for our sample. These stars were faint enough that they did not saturate the detector in ideal observing conditions. Known binaries stars were found using SIMBAD database\footnote{\url{http://simbad.u-strasbg.fr/simbad/}} and removed from the sample. The full target list, relevant photometry, and spectral types are shown in Table \ref{table::obslog}. The spectrum of BD --17 491 is shown in Fig. \ref{fig:spectra} to demonstrate the absorption features in our spectral range.

The observations were obtained at the Gemini South telescope using the Phoenix instrument \citep{hinkle_et_al_1998} for the program GS-2016B-Q-77. The 4 pixel slit was used on Phoenix to achieve a resolution of $\sim$50,000. We used the blocking filter L2734 to observe echelle order 15 covering the wavelength range between 36950 $\AA$ - 37115 $\AA$. Target stars were nodded along the slit and observed in 'abba' position pairs for sky subtraction during data reduction. Stars were nodded at 3.5" to avoid contamination during sky-subtraction from the broadened profile of the star present during poor seeing conditions. A and B type stars were observed and the airmass differences between the telluric standard observations and target observations were less than 0.1.  
 
\begin{deluxetable*}{ c c c c c c c c}
%\tabletypesize{10pt} 
%\rotate
\tablewidth{0pt} 
\tabletypesize{\footnotesize}
\tablecaption{Summary of Phoenix Observations \label{table::obslog}} 
 \tablehead{\colhead{2MASS} & \colhead{Other ID} & \colhead{UT Date}& \colhead{J\tablenotemark{a}} & \colhead{K$_{s}$\tablenotemark{a}} & \colhead{W3\tablenotemark{b}} & \colhead{Spectral\tablenotemark{c}} & \colhead{S/N} \\
 \colhead{Number} & \colhead{} & \colhead{Observed} & \colhead{(Mag)} & \colhead{(Mag)} & \colhead{(Mag)}& \colhead{Type} & \colhead{}  }
\startdata
2MASS J00243149-0954040  & GN Cet & 2016 Dec 15 & 4.052  &  2.714 & 1.826 & M6 & 190\\
2MASS J00465746-4758522 & AH Phe & 2016 Dec 8 & 3.764 & 2.45 & 2.146 & M6 III &300 \\
2MASS J02323698-1643360  & BD-17 491 & 2016 Dec 8  &  3.673	&	2.366 & 1.844 &	M5 &230\\
2MASS J04195770-1843196 & AV Eri & 2016 Dec 11& 4.254 &  2.822 & 1.683 & M6.5 & 180 \\
2MASS J07042577-0957580 & BQ Mon  & 2016 Dec 10& 3.888   &   2.027 &1.278 & M7 &220\\
2MASS J07300768-0923169 & KO Mon & 2016 Dec 15 &   4.312 & 2.67 & 1.566& M6 &180\\
\enddata
\tablenotetext{a}{J and K$_{s}$ magnitudes from 2MASS \citep{skrutskie06}}
\tablenotetext{b}{W3 magnitudes from WISE \citep{wright10}}
\tablenotetext{c}{spectral types from the SIMBAD database}

\end{deluxetable*} 
 
Standard IR data reduction procedures were followed \citep{joyce1992}. Data reduction was performed using the IRAF software suite\footnote{IRAF is distributed by the National Optical Astronomy Observatory, which is operated by the Association of Universities for Research in Astronomy, Inc., under cooperative agreement with the National Science Foundation.} follwing the same procedures as \citet{maas16}. To summarize, the images were were trimmed, flat-fielded using a dark corrected flat field image, and sky-subtracted. The spectra were extracted, average combined, normalized, and corrected for telluric lines. The wavelength solution was derived using stellar lines in the spectra. The telluric lines were sparse in our spectral range (shown in Fig. 2 in \citealt{maas16}) and provided an inferior wavelength solution than the stellar lines in these M stars. 

\section{Measuring the Cl Isotope Ratio}
\label{sec::methodology}

First, equivalent widths of the two HCl features at 36985 $\AA$ (H$^{35}$Cl) and 37010 $\AA$ (H$^{37}$Cl) were measured using the deblending tool in \texttt{splot}. Both HCl lines have nearly identical excitation potentials and log gf values \citep{rothman13}, which are listed in \citet{maas16}. The chlorine isotope ratio was found in RZ Ari by taking the ratio of the equivalent widths of the H$^{35}$Cl line and the H$^{37}$Cl line \citep{maas16}. Both HCl features in that star had weak line strengths on the linear portion of the curve of growth. We compared our 6 stars to curve of growth models for the two HCl lines and found small deviations from the linear approximation. The curves of growth were created using MOOG (\citealt{sneden73}, v. 2014) and MARCS model atmospheres \citep{gustafsson}. The full line list used to create the model spectrum in Fig. 1 is listed in \citet{maas16}.

\subsection{Atmospheric Parameters}
\label{subsec:params}

Atmospheric parameters are needed to determine the position of the HCl features on the curve of growth (COG). The effective temperature and microturbulence most impacted the shape of the COG when generating models. Spectral types for our stars were determined from the SIMBAD database and are listed in Table \ref{table::obslog}, however, no stars have atmospheric parameters derived in the literature. We first determined that our stars are giants from two arguments. The J-K$_{s}$ colors for the stars are consistent with giants: for example the intrinsic J-K$_{s}$ for an M5 giant is 1.36 while a dwarf M5 star has 0.77 \citep{jian17}. Additionally, the Ca I lines observed between 36960 - 36975 $\AA$ are broadened at higher gravities. The Ca I lines for the stars in our sample are similar to BD --17 491, shown in Fig. \ref{fig:spectra}, and are consistent with the spectra of giants. 

Temperatures were derived using the J-W3 color with J-band photometry from 2MASS \citep{skrutskie06} and the the W3 band from WISE \citep{wright10}. The temperature-color relation from \citet{jian17} was used and the temperature derived for each star is listed in Table \ref{table::params}. This temperature-color relation is calibrated for giants between 3650 K $<$ T$_{eff}$ $<$ 5100 K and so the relation was extrapolated to determine the temperatures for our stars. The spectral energy distribution was constructed for each star using photometry from 2MASS \citep{skrutskie06}, WISE \citep{wright10}, and IRAS \citep{beichman88}. The Rayleigh-Jeans tail of the spectral energy distribution (SED) was compared to a blackbody function to determine if any of the sample stars show a significant infrared excess. For each star, a single scaled blackbody function fits the infrared portion of the curve except for deviations at $\sim$ 100 $\mu$m. 

The J and W3 bands were used due to the uncertainty on the K$_{s}$, W1, and W2 photometric measurements. The W1 and W2 Wise bands were saturated and the average W3 band uncertainty is 0.02 $\pm$ 0.007 mags. The average J-band magnitude error is 0.27 $\pm$ 0.03 for the sample and the $K_{s}$ magnitudes had uncertainties similar to the J band magnitudes. The cores of the star images are saturated in the 2MASS photometry and the large photometric error was estimated from the fit to the unsaturated portion of the 1-D radial profile fit \citep{skrutskie06}. A J magnitude range of $\pm$ 0.27 mags translates into a temperature difference of $\sim$ 150 K for the stars in our sample. Due to the uncertainty on the J band magnitude and the extrapolation of the temperature-color relation, an uncertainty of $\pm$ 200 K is appropriate for our derived temperatures. 

The microturbulence values ($\xi$) were derived by fitting COG models to the OH lines in our spectral range. OH line equivalent widths were measured using the deblending tool within \texttt{splot} in IRAF. The equivalent width values are listed in Table \ref{table::oh}. Uncertainties were estimated by measuring the equivalent width multiple times at different continuum levels. Empirical curves of growth were created for each star with a range of microturbulence values in steps of 0.1 km/s. The models were created at the temperature of the star, log g = 0.5, and at solar metallicity. COGs were created using MOOG and OH excitation potential and log gf values from \citet{brooke16}. The log gf values were tested by fitting the spectra of Arcturus and the Sun \citep{maas16}. 

\begin{deluxetable*}{ cccccc}
%\tabletypesize{10pt} 
%\rotate
\tablewidth{0pt} 
\tabletypesize{\small}
\tablecaption{Atmospheric Parameters and Cl Isotope Ratios \label{table::params}} 
 \tablehead{\colhead{Star Name} & \colhead{Temperature} & \colhead{$\xi$} & \colhead{H$^{35}$Cl EW} & \colhead{H$^{37}$Cl EW} & \colhead{$^{35}$Cl/$^{37}$Cl} \\
 \colhead{} & \colhead{(K)} & \colhead{(km/s) } & \colhead{(m$\AA$)} & \colhead{(m$\AA$)} & \colhead{}}
\startdata
GN Cet & 3102 &	2.5 $\pm$ 0.2 & 165 $\pm$ 10 & 106 $\pm$ 8 & 1.76 $\pm$ 0.17 \\
AH Phe & 3516 & 2.5 $\pm$ 0.2 & 122 $\pm$ 11 & 63 $\pm$ 8 & 2.20 $\pm$ 0.30 \\
BD -17 491 & 3355 & 2.3 $\pm$ 0.2 &  137 $\pm$ 9 &  66 $\pm$ 8 & 2.42 $\pm$ 0.30 \\
AV Eri & 2921 & 1.4 $\pm$ 0.3 & 161 $\pm$ 12 & 62 $\pm$ 8 & 3.42 $\pm$ 0.50 \\
%IRAS 06015-0357 & 3020 & 1.4 $\pm$ 0.3 & 183 $\pm$ 10 & 83 $\pm$ 8 & 3.03 $\pm$ 0.34 \\ 
BQ Mon & 2902 & 1.6 $\pm$ 0.2 & 185 $\pm$ 11 & 81 $\pm$ 6 & 2.92 $\pm$ 0.31 \\
KO Mon & 2838 & 2.4 $\pm$ 0.2 & 243 $\pm$ 19 & 98 $\pm$ 9 & 3.22 $\pm$ 0.42 \\
RZ Ari\tablenotemark{a} & 3340 & 2.54 $\pm$ 0.15 & 81 $\pm$ 6 & 36 $\pm$ 6 & 2.2 $\pm$ 0.4 \\
\enddata
\tablenotetext{a}{Temperature from \citet{mcdonald12}; microturbulence from \citet{tsuji08}; Cl isotope ratio from \citet{maas16}}
%\tablenotetext{b}{J magnitudes from 2MASS \citep{skrutskie06}}
%\tablecomments{Telescopes (1) Gemini South Telescope;\\ (2) KPNO Mayall 4m Telescope}
\end{deluxetable*}

\begin{deluxetable*}{ lcccccc}
%\tabletypesize{10pt} 
%\rotate
\tablewidth{0pt} 
\tabletypesize{\scriptsize}
\tablecaption{OH Equivalent Width Measurements \label{table::oh}} 
 \tablehead{\colhead{Wavelength} & \colhead{GN Cet} & \colhead{AH Phe} & \colhead{BD -17 491} & \colhead{AV Eri} & \colhead{BQ Mon} & \colhead{KO Mon}   \\
 \colhead{($\AA$)} & \colhead{EW (m$\AA$)} & \colhead{EW (m$\AA$)}   & \colhead{EW (m$\AA$)}  & \colhead{EW (m$\AA$)} & \colhead{EW (m$\AA$)} & \colhead{EW (m$\AA$)}     }
\startdata
              36954.896 &      440 $\pm$ 30 &    468 $\pm$ 23 &    450 $\pm$ 32    &   441 $\pm$ 26 &     476 $\pm$ 43        &   502 $\pm$ 38  \\
              36979.19  &       74 $\pm$ 15 &     94 $\pm$ 20 &     81 $\pm$ 18    &   113 $\pm$ 18 &     120 $\pm$ 27        &  85 $\pm$ 15    \\
              36981.688 &      420 $\pm$ 21 &    438 $\pm$ 29 &    417 $\pm$ 19    &   399 $\pm$ 27 &     465 $\pm$ 31        &   422 $\pm$ 23 \\
              36998.379 &      527 $\pm$ 14 &    489 $\pm$ 27 &    491 $\pm$ 33    &   441 $\pm$ 33 &       491 $\pm$ 32        &   513 $\pm$ 28 \\
              37034.102 &      624 $\pm$ 46 &    675 $\pm$ 33 &      618 $\pm$ 27    &   575 $\pm$ 33 &     637 $\pm$ 34        &   674 $\pm$ 43   \\
              37050.462 &      655 $\pm$ 39 &    662 $\pm$ 28 &      642 $\pm$ 53    &   601 $\pm$ 24 &     628 $\pm$ 44        &   730 $\pm$ 45 \\
              37055.419 &      381 $\pm$ 40 &    338 $\pm$ 42 &      334 $\pm$ 51    &   316 $\pm$ 33 &     318 $\pm$ 40        &   356 $\pm$ 42 \\
              37059.961 &      724 $\pm$ 36 &    643 $\pm$ 55 &    599 $\pm$ 18    &   648 $\pm$ 34 &   637 $\pm$ 47        &   835 $\pm$ 48   \\
              37063.634 &      382 $\pm$ 37 &    384 $\pm$ 17 &    346 $\pm$ 25    &   309 $\pm$ 52 &     353 $\pm$ 14        &   392 $\pm$ 40 \\
\enddata
%\tablenotetext{a}{spectral types from the SIMBAD database}
%\tablenotetext{b}{J magnitudes from 2MASS \citep{skrutskie06}}
%\tablecomments{Telescopes (1) Gemini South Telescope;\\ (2) KPNO Mayall 4m Telescope}
\end{deluxetable*}

\begin{figure*}[t!]
	\centering
	\includegraphics[trim=0cm 0cm 0cm 0cm, scale=.35]{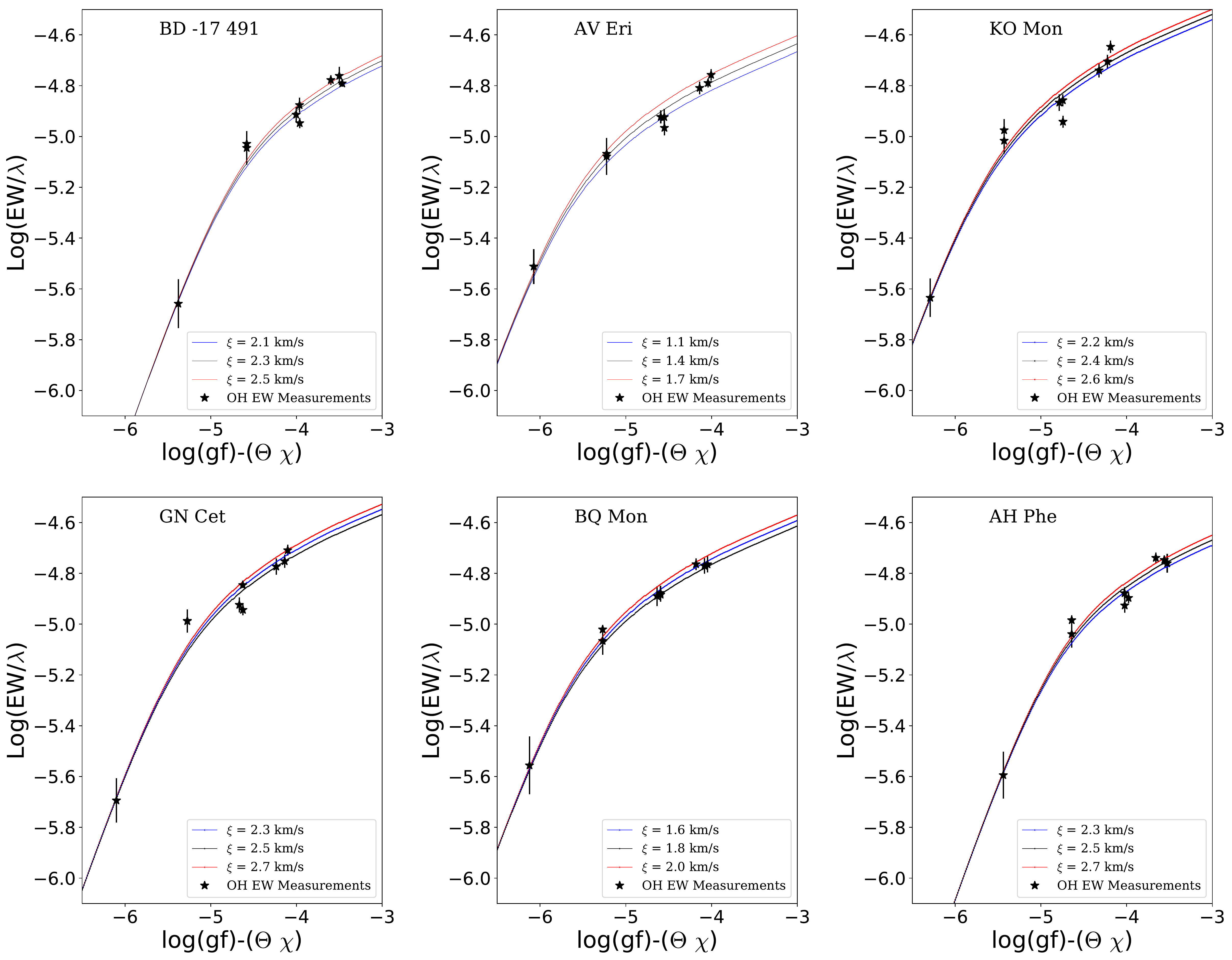}
	\caption{An empirical curve of growth for the OH lines measured for our sample. The x-axis is the log(gf) - ($\Theta$ $\chi$ ): where $\Theta$=5040/T(K) and $\chi$ is the excitation potential. The y-axis represents the reduced equivalent width. Three model curve of growths were created for three different microturbulence values ($\xi$). These models are fit to the measured reduced equivalent widths to determine the microturbulence for the star.\label{fig:microturb}}
	\end{figure*}

\begin{figure*}[t!]
	\centering
	\includegraphics[trim=0cm 0cm 0cm 0cm, scale=.30]{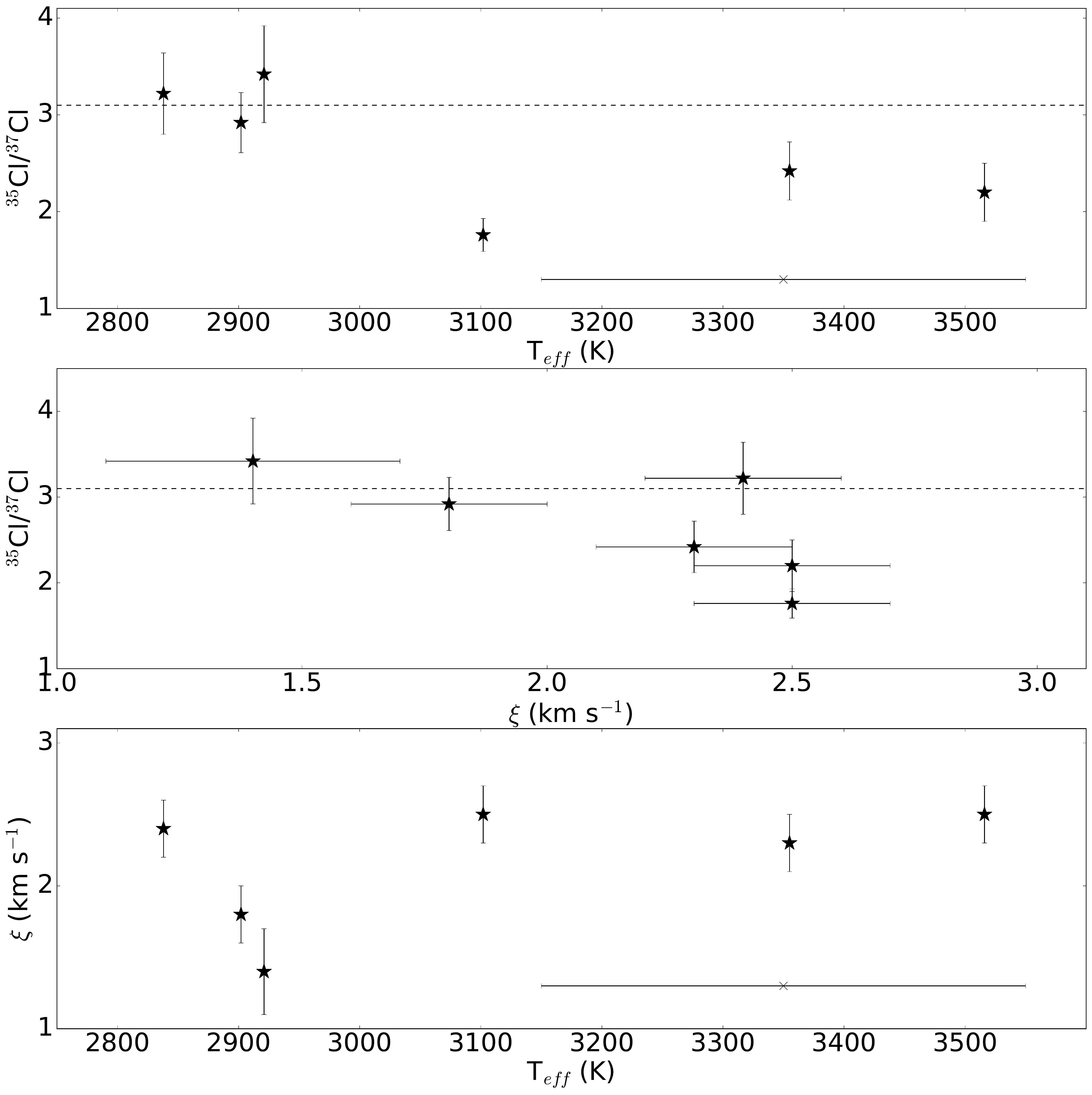}
	\caption{Three plots showing the relationship between the derived atmospheric parameters, effective temperature and microturbulence and the measured Cl isotope abundances in our sample of stars. The top plot shows isotope ratios versus temperature, the middle plot shows isotope ratios versus microturbulence, and finally the bottom plot shows microturbulence versus temperature. The uncertainty on the temperature is estimated to be 200 K for each of our stars, shown by the error bar in the top and bottom plots. \label{fig:results}  }
	\end{figure*}

Each empirical COG was shifted until the weak 36979 $\AA$ OH line fell on the model COG when fitting the data; this OH line approximately falls on the linear portion of the COG. Multiple models with different microturublence values, in steps of 0.1 km s$^{-1}$ were fit to the data. The model that resulted in the lowest $\chi^{2}$ was chosen as the best value. A Monte-Carlo simulation was performed to determine the error on the derived microturbulence using this method. The OH equivalent widths for each line were adjusted randomly for each iteration from a Gaussian distribution with a mean of the measured value for a line and a standard deviation of the measurement error. The process was repeated for 10,000 iterations and the uncertainty on the microturbulence was set at the 5 $\%$ and 95 $\%$ median values due to the large step sizes in microturbulence. The best fits for the six stars in the sample are shown in Figure \ref{fig:microturb}. We also compared the dependence of the derived temperature on microturbulence. No strong correlation between effective temperature and microturbulence was seen for our sample of stars, as shown in Fig. \ref{fig:results}. 

To test our method, we determined the microturbulence of RZ Ari since this star has an T$_{eff}$ of 3340 K, similar to our sample of M giants. The observations for this star are detailed in \citet{maas16}. The microturbulence from measurements of the OH lines lead to value of $\xi$ = 2.1$^{+0.3}_{-0.2}$ km s$^{-1}$. Our microturbulence measurement is compatible within the uncertainties with the microturbulence from \citet{tsuji08}, listed in Table \ref{fig:results}, and derived from weak molecular lines.   
 
Our derived microturbulence values are also consistent with values found in other evolved M stars. Microturbulences varied between 1.5 - 2.6 km s$^{-1}$ for carbon stars in the sample of \citet{lambert86}. Also, \citet{lambert86} noted that stars with the lowest microturbulence values, between 1.5 - 1.8 km s$^{-1}$ have anomalous features in their spectra, such as a Mira variable (R Lep), wavelength dependent CO lines abundances (R Scl), and strong CO lines with low abundances (V Hya). Without spectra covering larger wavelength ranges we cannot determine if these anomalies are present in the stars we observed with low microturbulence values. Other studies find microturbulence values for red giants between 2.1 - 3.2 km s$^{-1}$ \citep{cunha07}. 

Surface gravity was assumed to be log g = 0.5; varying the parameter by $\pm$ 0.5 created negligible corrections on the equivalent width, similar to the results of \citet{maas16}. We also assumed stars in our sample to have solar metallicities. Metal-poor stars with ([Fe/H] $\lesssim$ --1) likely could not reach the low effective temperatures found in our sample.

\subsection{Spectral Synthesis}
\label{subsec::synthesis}

To ensure blends with weak CN/CH lines did not affect the M giants in our sample, we fit synthetic spectra to the stars. The line-list was replicated from \citet{maas16}, which included OH lines from \citet{brooke16}, CN line features from \citet{brooke14} and accessed via the Kurucz database\footnote{\texttt{kurucz.harvard.edu/molecules}}, NH lines from \citet{brookenh14,brookenh15}, CH lines from \citet{masseron14}, and HCl line information from \citet{rothman13}. The synthetic spectra were generated using \texttt{pymoogi}\footnote{\url{https://github.com/madamow}}, a python wrapper that runs MOOG v. 2014 \citep{sneden73}. A plot of the synthetic spectrum for the M star BD --17 491 is  shown in Figure \ref{fig:spectra}. Atmospheric parameters for BD --17 491 are described in Section \ref{subsec:params} with the temperatures and microturbulent velocities listed in Table \ref{table::params}. The abundances were adjusted to provide the best fit to the spectrum for line identification. We find that in stars with C/O $<$ 1, OH lines dominate and CH/CN are weak or negligible. We therefore can measure HCl equivalent widths without blends in spectral type M stars in our sample. 

\subsection{Cl Isotope Ratios}

Model COGs for HCl were created using the effective temperatures and microturbulence values listed in Table \ref{table::params}, an assumed log g value of 0.5, and solar metallicities. The Cl isotope ratio derived from the ratio of equivalent width was adjusted by the distance between the linear approximation and the model curve of growth. Without the correction factor, larger equivalent widths would incorrectly represent smaller abundance. The corrections are most pronounced for the H$^{35}$Cl line due to its larger line strength. The final chlorine isotope ratio was calculated by taking a ratio of the equivalent widths with corrections for deviations from non-linearity. The results can be found in Table \ref{table::params}. 

The dependence of the Cl isotope ratios on photometric effective temperature and microturbulence is explored in Figure \ref{fig:results}. From inspection there appears no strong relationship between the microturbulence and isotopic abundance.  We estimated the slope between the Cl isotope ratio and effective temperature with a Monte Carlo simulation. Simulated isotope ratios and temperatures were generated as a random Gaussian number with the measured ratio as the mean value and the uncertainty as the standard deviation. A linear fit was done for each set of measured isotopic abundances for 10,000 iterations. The final result was an average slope of --0.1 $\pm$ 4.0 (ratio / 100K). A systematic uncertainty underlying a relationship between T$_{eff}$ and Cl isotope ratio would induce artificial scatter in the data. However, we note the warmest stars have the weakest HCl lines and therefore the smallest corrections. The Cl isotope ratio measured in these stars are likely the least impacted by any existing systematic effects. The Cl isotope ratio for AH Phe for example is 2.2 $\pm$ 0.3 which still is above the predicted chemical evolution model and is $\sim$ 3$\sigma$ from the solar value. We do not find a significant slope between temperature and abundance given our uncertainties. More accurate effective temperatures and a larger sample are necessary to determine if any relationship exists between effective temperature and Cl isotope ratio.

We used the \texttt{abfind} driver in MOOG to estimate the abundance of $^{35}$Cl and  $^{37}$Cl from the equivalent width measurements. The isotope ratios derived from the abundances agree with the COG method however, the uncertainties on the abundances were significantly larger. We tested the star BD --17 491 and calculated the A($^{35}$Cl) and A($^{37}$Cl) abundance for the best model parameters, for $\delta$T = $\pm$ 200 K, $\delta$log g = $\pm$ 0.5, [Fe/H] at -1 and 0.5, $\delta \xi$ $\pm$ 0.2 km s$^{-1}$, and for the one sigma uncertainties on the equivalent width measurement. The abundances were derived independently and the average difference between the high and low atmospheric parameter were taken as the uncertainty. Each term was added in quadrature to determine the total uncertainty. For BD --17 491, we determined A($^{35}$Cl) =  5.00 $\pm$ 0.96 dex and A($^{37}$Cl) = 4.61 $\pm$ 1.06 dex. The metallicity and temperature were the main sources of uncertainties, similar to the findings of \citet{maas16}. Cl abundances were not determined for the rest of our sample because of the large uncertainties. However, the isotope ratio between the two abundances is accurate since errors in the atmospheric parameters will affect both the H$^{35}$Cl and H$^{37}$Cl lines equally. The systematic errors will be removed except for uncertainties on the shape of the curve of growth given in Table \ref{table::uncertainty}. 

\subsection{Uncertainties}
Uncertainties were estimated for both the atmospheric parameters chosen and the equivalent width measurements. Atmospheric models were created at the 1$\sigma$ level for the stars effective temperature and microturbulence. Additional models were created at a log g of one, a log g of zero, at an [Fe/H] at --1 and 0.5 dex. The uncertainty on the COG nonlinear correction was computed for each atmospheric parameter independently. The uncertainty from each atmospheric parameter and the fit were added in quadrature to determine the final uncertainties on the H$^{35}$Cl and H$^{37}$Cl corrected equivalent width measurements. The dominant term is the uncertainty on the H$^{37}$Cl equivalent width. For example, the uncertainty on the isotope ratio for GN Cet is lower because the H$^{37}$Cl features is relatively strong. The average uncertainties on the equivalent width corrections are shown in Table \ref{table::uncertainty}.

\begin{deluxetable}{ ccc}
%\tabletypesize{10pt} 
%\rotate
\tablewidth{0pt} 
\tabletypesize{\small}
\tablecaption{Average Uncertainty on  Non-Linearity Correction from Atmospheric Parameters \label{table::uncertainty}} 
 \tablehead{\colhead{Parameter} & \colhead{H$^{35}$Cl} & \colhead{H$^{35}$Cl} \\
 \colhead{} & \colhead{(m$\AA$)} & \colhead{(m$\AA$) } }
\startdata
T$_{eff}$ &	11 & 2 \\
log g  & 4 & 1 \\
$[Fe/H]$ & 5 & 1 \\
$\xi$  & 7 & 1 \\
\enddata
\end{deluxetable}

\section{Discussion}

\label{sec::discussion}
\subsection{Chlorine Isotopologue Nucleosynthesis}
\label{subsec::isotope}
Our Cl isotope ratio measurements allow us to explore multiple important questions about the nucleosynethsis of Cl. First, is the Cl isotope ratio in the solar system consistent with our observed isotopic abundances? The meteoric solar Cl isotope ratio falls within the range of measurements in our sample, from 1.76 - 3.42, although it is at the high end.

Next, does our sample agree with the predicted Cl isotope ratio for the solar neighborhood? The chemical evolution model of \citet{kobayashi11} predicts Cl isotope ratios in the solar neighborhood of $^{35}$Cl/$^{37}$Cl = 1.94 at an [Fe/H] of --0.5 and a Cl isotope ratio of $^{35}$Cl/$^{37}$Cl = 1.80 at solar metallicity. Our average Cl isotope ratio is $\sim$ 1.5$\sigma$ higher than this value and AH Phe, BQ Mon, and KO Mon in particular deviate by 2$\sigma$ - 3$\sigma$ from the model prediction for solar metallicity stars. The model prediction is also smaller than the solar system meteoric Cl isotope ratio of 3.13 \citep{lodders09}. The offset may be due to either underproduction of $^{35}$Cl, or an overproduction of $^{37}$Cl in the models. Yields from supernova models that include rotation have larger $^{35}$Cl/$^{37}$Cl ratios for the 25 M$_{\odot}$ progenitors \citep{chieffi13} and may help explain the discrepancy between chemical evolution models and measurements. Inclusion of the $\nu$ process, which produces $^{35}$Cl may also impact yields from supernovae. \citet{maas16} suggests that $^{35}$Cl abundances are larger than expected from chemical evolution model of \citet{kobayashi11}. 

Next, is the spread of Cl isotope ratios observed consistent with the measurement errors? The mean isotope ratio and standard deviation of our sample is 2.66 $\pm$ 0.58. Since the measurement precision varied from star to star, we also calculated a weighted mean of $^{35}$Cl/$^{37}$Cl = 2.27 $\pm$ 0.11. 

We performed a Monte Carlo simulation to determine if the standard deviation from our six stars is consistent with scatter around a common average Cl isotope ratio or if the scatter in our data reflects differences in the surface composition of our stars. Six random numbers were generated using a Gaussian distribution with a mean equal to the weighted average of our sample and standard deviations derived for our sample stars. The standard deviation was measured for the six values. We continued measuring standard deviations for 50,000 sets of measurements and found that 98.8 $\%$ of the trials had standard deviations less than 0.58. The mean Cl isotope ratio standard deviation from the simulation was 0.30. The observed large standard deviation implies that some portion of the scatter in the Cl isotopic ratio seen in our sample reflects differences in surface abundance and the chemical history of the stars. 

The Gaia DR 1 did not include extremely blue or red stars which may be why no parallax measurements are available for our sample \citep{gaia1}.  Additionally, all stars but BD --17 491 are identified as long period variables in the SIMBAD database. AH Phe is the only star classified as luminosity class III and we estimate the distance to this star based on this classification. First, we estimate a luminosity of log(L/L$_{\odot}$) = 3.1 for a 1.5 $M_{\odot}$, Z = 0.017 star at T$_{eff}$ $\sim$ 3500 using stellar evolutionary tracks from \citet{bertelli08,bertelli09}. We use the K$_{s}$ measurement from Table \ref{table::obslog} and a bolemetric correction of 3 from \citet{bessell98} to determine the distance modulus. We used the color transformation relations of \citet{johnson05} to convert the K$_{s}$ measurements to the Johnson K system. We estimate a distance of $\sim$ 520 pc to AH Phe\footnote{Gaia DR2 was released after the manuscript was submitted and found a parallax of (1.68 $\pm$ 0.08) mas for AH Phe \citep{gaiadr2}. The distance to AH Phe is therefore (595 $\pm$ 28) pc}.

\begin{deluxetable*}{llllc}
%\tabletypesize{10pt} 
%\rotate
\tablewidth{0pt} 
\tabletypesize{\small}
\tablecaption{Cl Isotope Ratios Measured in the ISM and Stars \label{table::ism_ratio}} 
 \tablehead{\colhead{Object} & \colhead{Object Type} & \colhead{Cl Ratio} & \colhead{Feature} & \colhead{Reference} }
\startdata
IRC+10216 & Carbon Star &	2.3 $\pm$ 0.5 & NaCl, AlCl  & (1) \\
IRC+10216 & Carbon Star& 3.1 $\pm$ 0.6 & NaCl, AlCl, KCl & (2) \\
IRC+10216 & Carbon Star& 2.3 $\pm$ 0.24 & NaCl, AlCl & (3) \\
IRC+10216 & Carbon Star & 3.3 $\pm$ 0.3 & HCl & (4) \\ 
CRL 2688 & Post-AGB (Pre-PN) Star & 2.1 $\pm$ 0.8 & NaCl &  (5) \\
Orion A & Giant Molecular Cloud (OMC-1)  & $\sim$ 4 - 6 & HCl & (6) \\
Multiple Objects & Multiple Objects\tablenotemark{a} & 1 - 5 & HCl & (7) \\
NGC6334I & Molecular Gas in Embedded Star Cluster/SFR\tablenotemark{b} & 2.7 & H$_{2}$Cl$^{+}$ & (8) \\
Sgr A, W31C & Foreground Molecular Clouds towards SgrA, W31C & $\sim$ 2-4& H$_{2}$Cl$^{+}$ & (9) \\
PKS 1830-211 & Foreground Absorbed towards Lensed Blazar & 3.1$^{+0.3}_{-0.2}$ & H$_{2}$Cl$^{+}$ & (10) \\
W49N & Diffuse Foreground Absorbers towards H II Region W49N & 3.50$^{+0.21}_{-0.62}$ & H$_{2}$Cl$^{+}$ & (11) \\
W3 & Molecular Cloud toward HII region W3  & 2.1 $\pm$ 0.5 & HCl & (12) \\
CRL 2136 & Molecular Cloud associated with SFR CRL 2136 & 2.3 $\pm$ 0.4 & HCl & (13) \\
W31C &  Absorber towards SFR W31C & 2.9 & HCl & (14) \\
OMC-2-FIR-4 & Proto-Stellar Core in Orion Nebula& 3.2 $\pm$ 0.1 & HCl & (15) \\
W31C & Absorber towards SFR W31C & 2.1 $\pm$ 1.5 & HCl$^{+}$ & (16) \\ 
RZ Ari & Red Giant Star & 2.2 $\pm$ 0.4 & HCl & (17) \\
Multiple Stars & Red Giant Stars & 1.76 - 3.42 & HCl & This Work \\
\enddata
\tablenotetext{a}{Survey included the circumstellar envelopes of evolved stars and molecular clouds in along the lines of sight towards star forming regions.}
\tablenotetext{b}{Star Forming Region (SFR)}
\tablecomments{References: (1) \citealt{cernicharo87}; (2) \citealt{cernicharo00}; (3) \citealt{kahane00}; (4) \citealt{agundez11}; (5) \citealt{highberger03}; (6) \citealt{salez96}; (7) \citealt{peng10}; (8) \citealt{lis09}; (9) \citealt{neufeld12}; (10) \citealt{muller14}; (11) \citealt{neufeld15}; (12) \citealt{cernicharo10}; (13) \citealt{goto13}; (14) \citealt{monje13} ;(15) \citealt{kama15}; (16) \citealt{luca12} ;(17) \citealt{maas16}}
\end{deluxetable*}

	\begin{figure*}[t!]
	\centering
	\includegraphics[trim=0cm 0cm 0cm 0cm, scale=.37]{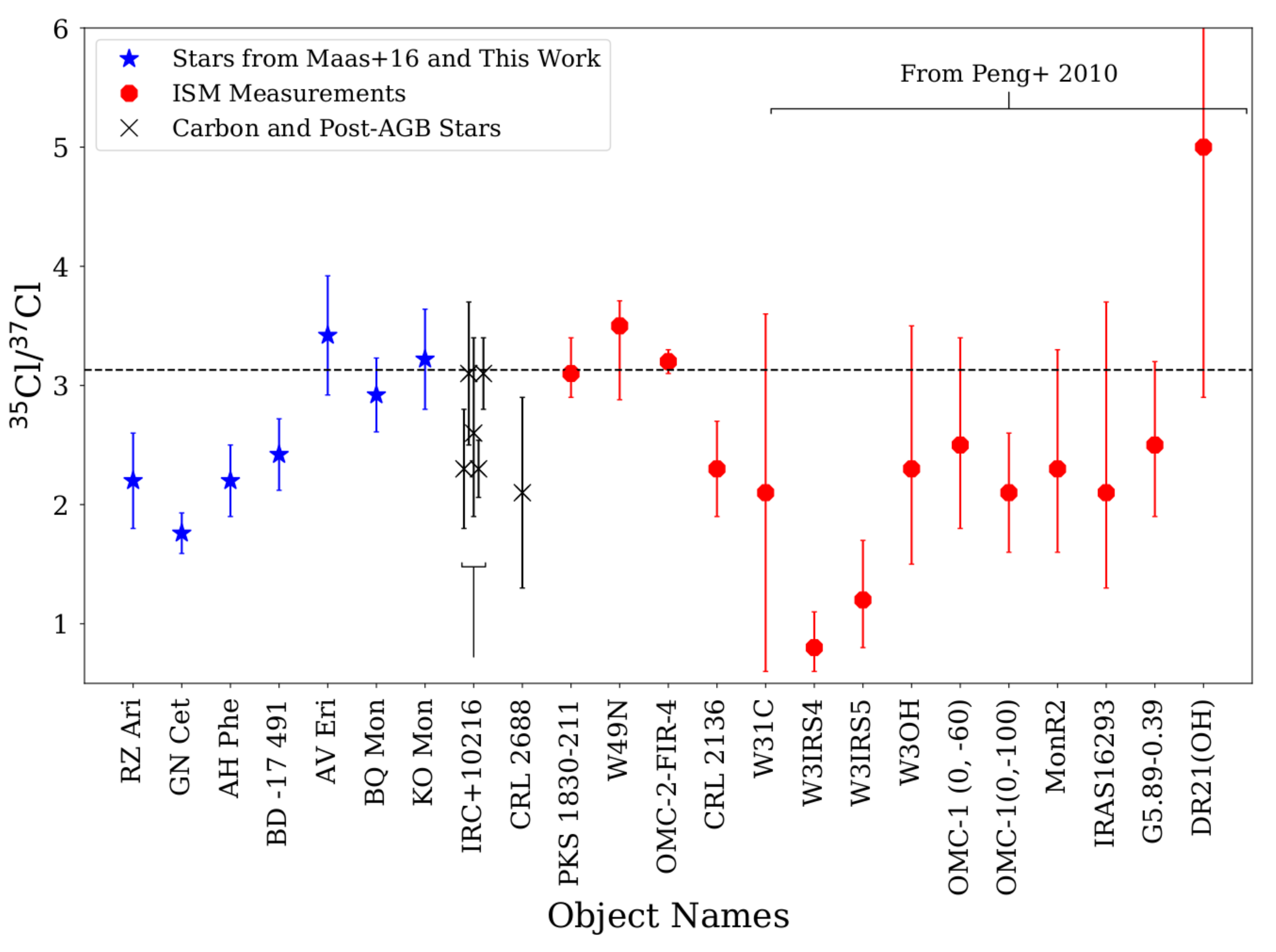}
	\caption{Cl isotope ratios from Table \ref{table::ism_ratio} plotted as a function of object name and Cl isotope ratio. The dashed line represents the solar system meteoric Cl isotope ratio. Blue star represent Cl isotope ratios measured in M giants, red octogons are ISM measurements, and black crosses are measurements in the circustellar envelopes of carbon stars and post-AGB stars.  \label{fig:ism}  }
	\end{figure*}

\subsection{Comparisons to Cl Isotope Ratios in ISM}
\label{subsec::ism}
We compare our measurements of $^{35}$Cl/$^{37}$Cl to those made in star forming regions, molecular clouds, and the circumstellar envelopes of evolved stars, summarized in Table \ref{table::ism_ratio}. We find the Cl isotope ratios in stars are consistent with the range measured in the ISM; results from both studies with uncertainties are displayed in Figure \ref{fig:ism}. The measurements in the ISM are consistent with the solar system value, noted by \citet{muller14,neufeld15}, including a $^{35}$Cl/$^{37}$Cl = 3.1$^{+0.3}_{-0.2}$ measurement at a redshift of 0.89 \citep{muller14}.  Finally, the HCl survey of \citet{peng10} found a range of Cl isotopic ratios between 1-5. We find a smaller range between 1.76 $\leq$ $^{35}$Cl/$^{37}$Cl $\leq$ 3.42 of Cl isotope in stars.  

Local explosive nucleosynthesis events may partially explain the spread in isotope ratios both in the ISM \citep{salez96,peng10} and in our sample of stars. Studies of dust grains from the winds of evolved stars suggests the disk of the Galaxy well mixed and placed an upper limit on the spread of Si and Ti isotopes due to heterogeneous mixing in the disk at 40$\%$ \citep{nittler05}. A Monte Carlo simulation found the spread in [Fe/H] from heterogeneous mixing was expected to be between 0.022 to 0.03 dex depending on the yields adopted \citep{nittler05}. Assuming solar metallicity, these variations correspond to 5$\%$ - 7$\%$ variations in the ISM. For our average Cl isotope ratio of 2.66, 7$\%$ variations corresponds to a range of 0.19. The Monte Carlo simulation performed in section \ref{subsec::isotope} found an expected mean expected standard deviation of 0.30 dex, therefore the heterogeneous mixing in the ISM may explain some of the scatter in our data beyond the measurement uncertainties. Metallicity variations may also play a role in the dispersion seen in the ISM and in our sample of stars. The chemical evolution model of \citet{kobayashi11} predicts at [Fe/H] = 0, a $^{35}$Cl/$^{37}$Cl = 1.80, at an [Fe/H] = --0.5 a $^{35}$Cl/$^{37}$Cl = 1.94, and finally the prediction at [Fe/H] = --1.1 is $^{35}$Cl/$^{37}$Cl = 2.62.

Core collapse supernova are thought to be the dominant source of chlorine production and the isotope ratio varies depending on the progenitor star mass, metallicity, and explosion energy. For example, the larger the progenitor star at high metallicity, the lower the lower the Cl isotope ratio, as shown in Table \ref{table::yields}. Low mass progenitors may help explain observations of high isotope ratios (e.g. AV Eri with $^{35}$Cl/$^{37}$Cl = 3.42 $\pm$ 0.50) and visa-versa. Also, a 25 M$_{\odot}$, Z = 0.02 supernova with an explosion energy of E = 10$^{51}$ erg yields a Cl ratio of $^{35}$Cl/$^{37}$Cl $\sim$ 1.2, the same model parameters with E = 10$^{52}$ erg produces an isotope ratio of $^{35}$Cl/$^{37}$Cl = 1.75 \citep{kobayashi11}. The Cl isotope ratio from supernova typically increases as the metallicity of the progenitor star decreases. For example, a 25 $M_{\odot}$, Z = 0.004 supernova would have yields of $^{35}$Cl/$^{37}$Cl = 1.75 \citep{kobayashi06}, compared to a yield a ratio of 1.2 for a Z = 0.02 (see Table \ref{table::yields}).

Studies of the circumstellar material around stars suggested the s-process is creating $^{37}$Cl in AGB stars and reducing the Cl isotope ratio. Low Cl isotope ratios were measured in the circumstellar envelopes of IRC+10216 \citep{kahane00} and CRL 2688 \citep{highberger03} (listed in Table \ref{table::ism_ratio}). Both papers suggested s-process element enhancement in the stars may have caused Cl isotope ratios below the solar system values. However, other studies have found Cl isotope ratios consistent with the solar system value in IRC+10216 and the errors are significant for Cl 2688. Additionally, models only predict modest enhancement of $^{37}$Cl in AGB stars \citep{cristallo15,karakas16}. Additionally, our sample of stars were classified through analysis of low resolution spectroscopy and did not show signs of s-process enhancement indicative of an S star \citep{hansen75,houk78,kwok97}. 
 
\section{Summary}
\label{sec::conclusion}
We measured Cl isotope ratios in six M giants using spectra obtained from Phoenix on Gemini South. Measurements of Cl isotope ratios were performed using effective temperatures derived from J - W3 color and microturbulences from OH equivalent widths and curve of growth models. Cl isotope ratios were found by comparing equivalent widths from HCl 1-0 P8 features. Non-linearity corrections between the equivalent width and abundance were made using curve of growth models. From our Cl isotope ratios we have derived the following results:

\begin{enumerate}
\item{We find an average Cl isotope ratio and standard deviation of 2.66 $\pm$ 0.58 for our sample of stars. A Monte Carlo simulation suggests the scatter in our measurements reflects differences in surface abundances with over 2$\sigma$ confidence.}
\item{We find a range of Cl isotope ratios between 1.76 and 3.42 in our sample of stars. The solar system isotope ratio of 3.13 \citep{lodders09} is consistent with this range of measurements.}
\item{Our average Cl isotope ratio is higher than the predicted value of 1.80 for the solar neighborhood at solar metallicity \citep{kobayashi11} }
\item{The Cl isotope ratios measured in our sample of stars are consistent with those found in the ISM. The range in Cl isotope ratios may partially be due to different nucleosynthesis events enriching different portions of the ISM and from metallicity variations.}
\end{enumerate}

\section{Acknowledgements}
This work is based on observations obtained at the Gemini Observatory, which is operated by the Association of Universities for Research in Astronomy, Inc., under a cooperative agreement with the NSF on behalf of the Gemini partnership: the National Science Foundation (United States), the National Research Council (Canada), CONICYT (Chile), Ministerio de Ciencia, Tecnolog\'{i}a e Innovaci\'{o}n Productiva (Argentina), and Minist\'{e}rio da Ci\^{e}ncia, Tecnologia e Inova\c{c}\~{a}o (Brazil). The Gemini observations were done under proposal ID GS-2016B-Q-77. We thank Steve Margheim for his assistance with the Gemini South Telescope observing run. This research has made use of the NASA Astrophysics Data System Bibliographic Services, the Kurucz atomic line database operated by the Center for Astrophysics. This research has made use of the SIMBAD database, operated at CDS, Strasbourg, France. This publication makes use of data products from the Two Micron All Sky Survey, which is a joint project of the University of Massachusetts and the Infrared Processing and Analysis Center/California Institute of Technology, funded by the National Aeronautics and Space Administration and the National Science Foundation. This publication makes use of data products from the Wide-field Infrared Survey Explorer, which is a joint project of the University of California, Los Angeles, and the Jet Propulsion Laboratory/California Institute of Technology, funded by the National Aeronautics and Space Administration. We thank the anonymous referee for their thoughtful comments and suggestions on the manuscript. We thank Eric Ost for implementing the model atmosphere interpolation code. C. A. P. acknowledges the generosity of the Kirkwood Research Fund at Indiana University. 

\software{\texttt{IRAF}, \texttt{MOOG} (v2017; \citealt{sneden73}),  \texttt{pymoogi}, \texttt{scipy} \citep{jones01}, \texttt{numpy} \citep{walt11}, \texttt{matplotlib} \citep{hunter07}}

%\FloatBarrier

\FloatBarrier


\begin{thebibliography}{}

\bibitem[Ag{\'u}ndez et al.(2011)]{agundez11} 
Ag{\'u}ndez, M., Cernicharo, J., Waters, L.~B.~F.~M., et al.\ 2011, \aap, 533, L6 

\bibitem[Asplund et al.(2009)]{asplund09} %Solar Abundances
Asplund, M., Grevesse, N., Sauval, A.~J., \& Scott, P.\ 2009, \araa, 47, 481 

\bibitem[Beichman et al.(1988)]{beichman88}
 Beichman, C.~A., Neugebauer, G., Habing, H.~J., Clegg, P.~E., \& Chester, T.~J.\ 1988, Infrared astronomical satellite (IRAS) catalogs and atlases.~Volume 1: Explanatory supplement, 1,  

\bibitem[Bertelli et al.(2008)]{bertelli08} %stellar evolution tracts
Bertelli, G., Girardi, L., Marigo, P., \& Nasi, E.\ 2008, \aap, 484, 815 

\bibitem[Bertelli et al.(2009)]{bertelli09}  %stellar evolution tracts
Bertelli, G., Nasi, E., Girardi, L., \& Marigo, P.\ 2009, \aap, 508, 355

\bibitem[Bessell et al.(1998)]{bessell98} %bol Corrections
Bessell, M.~S., Castelli, F., \& Plez, B.\ 1998, \aap, 333, 231

\bibitem[Brooke et al.(2014)]{brooke14} %CN line information
Brooke, J.~S.~A., Ram, R.~S., Western, C.~M., et al.\ 2014, \apjs, 210, 23 

\bibitem[Brooke et al.(2014)]{brookenh14}%NH line info
Brooke J. S. A., Bernath P. F., Western C. M., van Hermert M. C., \& Groenenboom G. C., 2014, J. Chem. Phys., 141, 054310

\bibitem[Brooke et al.(2015)]{brookenh15}%NH line info
Brooke J. S. A., Bernath P. F., \& Western C. M., 2015, J. Chem. Phys., 143, 026101

\bibitem[Brooke et al.(2016)]{brooke16}%OH line info
Brooke J. S.A., Bernath P. F., Western C. M., et al., 2016, J. Quant. Spec. Rad. Trans., 168, 142

\bibitem[Cernicharo \& Guelin(1987)]{cernicharo87} 
Cernicharo, J., \& Guelin, M.\ 1987, \aap, 183, L10 

\bibitem[Cernicharo et al.(2000)]{cernicharo00} 
Cernicharo, J., Gu{\'e}lin, M., \& Kahane, C.\ 2000, \aaps, 142, 181 

\bibitem[Cernicharo et al.(2010)]{cernicharo10} 
Cernicharo, J., Goicoechea, J.~R., Daniel, F., et al.\ 2010, \aap, 518, L115 

\bibitem[Chieffi \& Limongi(2013)]{chieffi13} 
Chieffi, A., \& Limongi, M.\ 2013, \apj, 764, 21 

\bibitem[Codella et al.(2012)]{codella12} 
Codella, C., Ceccarelli, C., Bottinelli, S., et al.\ 2012, \apj, 744, 164 

\bibitem[Cristallo et al.(2015)]{cristallo15} %Fruity Database
Cristallo, S., Straniero, O., Piersanti, L., \& Gobrecht, D.\ 2015, \apjs, 219, 40

\bibitem[Cunha et al.(2007)]{cunha07} Cunha, K., Sellgren, K., Smith, V.~V., et al.\ 2007, \apj, 669, 1011 

\bibitem[De Luca et al.(2012)]{luca12} 
De Luca, M., Gupta, H., Neufeld, D., et al.\ 2012, \apjl, 751, L37 

\bibitem[Gaia Collaboration et al.(2016)]{gaia1} 
Gaia Collaboration, Prusti, T., de Bruijne, J.~H.~J., et al.\ 2016, \aap, 595, A1 

\bibitem[Gaia Collaboration et al.(2018)]{gaiadr2} 
Gaia Collaboration, Brown, A.~G.~A., Vallenari, A., et al.\ 2018, arXiv:1804.09365 

\bibitem[Goto et al.(2013)]{goto13}
 Goto, M., Usuda, T., Geballe, T.~R., et al.\ 2013, \aap, 558, L5 

\bibitem[Gustafsson et al.(2008)]{gustafsson} %MARCS models paper
Gustafsson, B., Edvardsson, B., Eriksson, K., et al.\ 2008, \aap, 486, 951

\bibitem[Hansen \& Blanco(1975)]{hansen75} 
Hansen, O.~L., \& Blanco, V.~M.\ 1975, \aj, 80, 1011 

\bibitem[Highberger et al.(2003)]{highberger03} 
Highberger, J.~L., Thomson, K.~J., Young, P.~A., Arnett, D., \& Ziurys, L.~M.\ 2003, \apj, 593, 393

\bibitem[Hinkle et al.(1995)]{hinkle95} %Arcturus Atlas
Hinkle, K., Wallace, L., \& Livingston, W.\ 1995, \pasp, 107, 1042 

\bibitem[Hinkle et al.(1998)]{hinkle_et_al_1998}
Hinkle, K. H., Cuberly, R., Gaughan, N., et al. 1998, Proc. SPIE, 3354, 810

\bibitem[Houdashelt et al.(2000)]{houdashelt00} 
Houdashelt, M.~L., Bell, R.~A., Sweigart, A.~V., \& Wing, R.~F.\ 2000, \aj, 119, 1424 

\bibitem[Houk(1978)]{houk78} 
Houk, N.\ 1978, Ann Arbor : Dept.~of Astronomy, University of Michigan : distributed by University Microfilms International, 1978-,  

\bibitem[Hunter(2007)]{hunter07}
Hunter, J. D. 2007, Computing in Science $\&$ Engineering, 9,
90. \url{http://scitation.aip.org/content/aip/journal/
cise/9/3/10.1109/MCSE.2007.55}

\bibitem[Jian et al.(2017)]{jian17} 
Jian, M., Gao, S., Zhao, H., \& Jiang, B.\ 2017, \aj, 153, 5 

\bibitem[Johnson et al.(2005)]{johnson05} %Convert Ks to K
Johnson, C.~I., Kraft, R.~P., Pilachowski, C.~A., et al.\ 2005, \pasp, 117, 1308 

\bibitem[Jones et al.(2001)]{jones01} %scipy
Jones, E., Oliphant, T., Peterson, P., et al. 2001, SciPy: 
Open source scientific tools for Python, , .\url{http://www.scipy.org/s}

\bibitem[Joyce(1992)]{joyce1992}
Joyce, R. R. 1992, in ASP Conf. Ser. 23, Astronomical CCD Observing and
Reduction Techniques, ed. S. Howell (San Francisco: ASP), 258

\bibitem[Kahane et al.(2000)]{kahane00}
 Kahane, C., Dufour, E., Busso, M., et al.\ 2000, \aap, 357, 669

\bibitem[Kama et al.(2015)]{kama15} 
Kama, M., Caux, E., L{\'o}pez-Sepulcre, A., et al.\ 2015, \aap, 574, A107

\bibitem[Karakas \& Lugaro(2016)]{karakas16} %AGB yields
Karakas, A.~I., \& Lugaro, M.\ 2016, \apj, 825, 26 

\bibitem[Keenan \& Boeshaar(1980)]{keenan80} % GG Pup Classification
Keenan, P.~C., \& Boeshaar, P.~C.\ 1980, \apjs, 43, 379 

\bibitem[Kobayashi et al.(2006)]{kobayashi06}
 Kobayashi, C., Umeda, H., Nomoto, K., Tominaga, N., \& Ohkubo, T.\ 2006, \apj, 653, 1145 

\bibitem[Kobayashi et al.(2011)]{kobayashi11} Kobayashi, C., Karakas, A.~I., \& Umeda, H.\ 2011, \mnras, 414, 3231 

\bibitem[Kwok et al.(1997)]{kwok97} 
Kwok, S., Volk, K., \& Bidelman, W.~P.\ 1997, \apjs, 112, 557 

\bibitem[Lambert et al.(1986)]{lambert86} 
Lambert, D.~L., Gustafsson, B., Eriksson, K., \& Hinkle, K.~H.\ 1986, \apjs, 62, 373

\bibitem[Leung \& Nomoto(2017)]{leung17} 
Leung, S.-C., \& Nomoto, K.\ 2017, arXiv:1710.04254 

\bibitem[Lis et al.(2010)]{lis09} 
Lis, D.~C., Pearson, J.~C., Neufeld, D.~A., et al.\ 2010, \aap, 521, L9 

\bibitem[Lodders et al.(2009)]{lodders09} %Chlorine meteorites
Lodders, K., Palme, H., \& Gail, H.-P.\ 2009, Landolt B{\"o}rnstein, 

\bibitem[Maas et al.(2016)]{maas16} Maas, Z.~G., Pilachowski, C.~A., \& Hinkle, K.\ 2016, \aj, 152, 196

\bibitem[MacConnell(1979)]{macconnell79} %Iras 06 Classification 
MacConnell, D.~J.\ 1979, \aaps, 38, 335

\bibitem[Masseron et al.(2014)]{masseron14} %ch lines 
Masseron, T., Plez, B., Van Eck, S., et al.\ 2014, \aap, 571, A47 

\bibitem[McDonald et al.(2012)]{mcdonald12} %temps
McDonald, I., Zijlstra, A.~A., \& Boyer, M.~L.\ 2012, \mnras, 427, 343

\bibitem[Monje et al.(2013)]{monje13} 
Monje, R.~R., Lis, D.~C., Roueff, E., et al.\ 2013, \apj, 767, 81 

\bibitem[Muller et al.(2014)]{muller14}
 Muller, S., Black, J.~H., Gu{\'e}lin, M., et al.\ 2014, \aap, 566, L6

\bibitem[Neufeld et al.(2012)]{neufeld12}
 Neufeld, D.~A., Roueff, E., Snell, R.~L., et al.\ 2012, \apj, 748, 37 

\bibitem[Neufeld et al.(2015)]{neufeld15} 
Neufeld, D.~A., Black, J.~H., Gerin, M., et al.\ 2015, \apj, 807, 54

\bibitem[Nittler(2005)]{nittler05} 
Nittler, L.~R.\ 2005, \apj, 618, 281

\bibitem[Nomoto et al.(2013)]{nomoto13}
Nomoto, K., Kobayashi, C., \& Tominaga, N.\ 2013, \araa, 51, 457 

\bibitem[Peng et al.(2010)]{peng10} 
Peng, R., Yoshida, H., Chamberlin, R.~A., et al.\ 2010, \apj, 723, 218

\bibitem[Pignatari et al.(2010)]{pignatari10} Pignatari, M., Gallino, R., Heil, M., et al.\ 2010, \apj, 710, 1557

\bibitem[Pignatari et al.(2016)]{pignatari16} 
Pignatari, M., Herwig, F., Hirschi, R., et al.\ 2016, \apjs, 225, 24

\bibitem[Pilachowski et al.(2017)]{pilachowski17} 
Pilachowski, C.~A., Hinkle, K.~H., Young, M.~D., et al.\ 2017, \pasp, 129, 024006 

\bibitem[Prantzos et al.(1990)]{prantzos90} Prantzos, N., Hashimoto, M., \& Nomoto, K.\ 1990, \aap, 234, 211

\bibitem[Rothman et al.(2013)]{rothman13} %Hitran Source
Rothman, L. S., Gordon, I. E., Babikov, Y., et al. 2013, JQSRT, 130, 4

\bibitem[Salez et al.(1996)]{salez96} %Orion Cl ratio 
 Salez, M., Frerking, M.~A., \& Langer, W.~D.\ 1996, \apj, 467, 708

\bibitem[Skrutskie et al.(2006)]{skrutskie06} 
Skrutskie, M.~F., Cutri, R.~M., Stiening, R., et al.\ 2006, \aj, 131, 1163 

\bibitem[Sneden(1973)]{sneden73} %MOOG
Sneden, C.\ 1973, \apj, 184, 839

\bibitem[Stephenson(1984)]{stephenson84} % Wy Pyx & CD-27 Classification
Stephenson, C.~B.\ 1984, Publications of the Warner \& Swasey Observatory, 3, 1 

\bibitem[Thielemann \& Arnett(1985)]{thielemann85} 
Thielemann, F.~K., \& Arnett, W.~D.\ 1985, \apj, 295, 604 

\bibitem[Travaglio et al.(2004)]{travaglio04} %Type Ia model
Travaglio, C., Hillebrandt, W., Reinecke, M., \& Thielemann, F.-K.\ 2004, \aap, 425, 1029

\bibitem[Tsuji(2000)]{tsuji00} 
Tsuji, T.\ 2000, \apj, 538, 801 

\bibitem[Tsuji(2008)]{tsuji08}
Tsuji, T.\ 2008, \aap, 489, 1271 

\bibitem[Wallace et al.(1993)]{wallace93} Wallace, L., Livingston, W.~C., \& Hinkle, K.\ 1993, An atlas of the solar photospheric spectrum in the region from 8900 to 13600 cm{$^{-1}$}(7350 to 11230 {\AA}), ~NSO technical report ; 93-001.~.~National Solar Observatory (U.S.), 

\bibitem[van der Walt et al.(2011)]{walt11}
van der Walt, S., Colbert, S. C., $\&$ Varoquaux, G. 2011,
Computing in Science $\&$ Engineering, 13, 22.
\url{http://scitation.aip.org/content/aip/journal/cise/13/2/10.1109/MCSE.2011.37}

\bibitem[Woosley et al.(1973)]{woosley73} 
Woosley, S.~E., Arnett, W.~D., \& Clayton, D.~D.\ 1973, \apjs, 26, 231

\bibitem[Woosley \& Weaver(1995)]{woosley95} Woosley, S.~E., \& Weaver, T.~A.\ 1995, \apjs, 101, 181 

\bibitem[Wright et al.(2010)]{wright10} %WISE
Wright, E.~L., Eisenhardt, P.~R.~M., Mainzer, A.~K., et al.\ 2010, \aj, 140, 1868-1881 

\end{thebibliography}
\end{document}